%% file: main.tex
\documentclass{article}
\usepackage[utf8]{inputenc}

\input{math_commands.tex}

\usepackage{hyperref}
\usepackage{url}
\usepackage{graphicx}
\usepackage{caption}
\usepackage{subcaption}
\usepackage{amsmath}
\usepackage{fullpage}

\title{TopTemp: Parsing Precipitate Structure from Temper Topology}

\author{Lara Kassab, Scott Howland, Henry Kvinge\thanks{Dr. Kvinge holds a joint appointment in the Department of Mathematics at the University of Washington.},  Keerti Sahithi Kappagantula, and Tegan Emerson\thanks{ Dr. Emerson holds joint appointments in the Department of Mathematics at Colorado State University and the Department of Mathematical Sciences at the University of Texas, El Paso.} \\
Pacific Northwest National Laboratory\\
Seattle, Washington 98109 \\
\texttt{\{lara.kassab, tegan.emerson\}@pnnl.gov} \\
}

\begin{document}

\maketitle

\begin{abstract}
Technological advances are in part enabled by the development of novel manufacturing processes that give rise to new materials or material property improvements. Development and evaluation of new manufacturing methodologies is labor-, time-, and resource-intensive expensive due to complex, poorly defined relationships between advanced manufacturing process parameters and the resulting microstructures. In this work, we present a topological representation of temper (heat-treatment) dependent material micro-structure, as captured by scanning electron microscopy, called TopTemp. We show that this topological representation is able to support temper classification of microstructures in a data limited setting, generalizes well to previously unseen samples, is robust to image perturbations, and captures domain interpretable features. The presented work outperforms conventional deep learning baselines and is a first step towards improving understanding of process parameters and resulting material properties.
\end{abstract}

\section{Introduction}
\label{sec:intro}

Material microstructure dictates a sample’s properties. Subject matter experts (SMEs) such as materials or manufacturing scientists and engineers use microstructures to help guide experimental design and inform selection of the relevant processing parameters. Material scientists seek to harness advances in deep learning in order to reduce the resource-intensiveness of this traditionally human-in-the-loop analysis. However, the viability of deep learning approaches is
reduced due to limited data and an accompanying preference for interpretability and explainability.

Evaluation of material microstructures can be performed through consideration of images derived from Scanning Electron Microscopy (SEM) amongst other modalities. Material scientists look at the relative size and distribution of structures present in the image such as grain size, grain boundary density, precipitate density, precipitate morphology and distribution. Tempering refers to heat treatments applied to metals and alloys to develop desirable microstructural features resulting in properties of interest. During tempering, a sample is exposed to specific temperatures for fixed lengths of time. The effects of tempering can be seen partially in SEM images and more distinctly higher resolution imaging techniques. It is well known in materials and manufacturing domains that there is a strong connection between the temper applied on a sample and its microstructure,  and the corresponding properties. The ability to capture explainable, relevant features tied to temper using machine-assisted methodologies may therefore form a viable base for exploring the more complex relationships between process conditions used to manufacture the sample and its properties. Even more useful is the ability to distinguish between samples with different temper using SEM images alone precluding the need for expensive imaging. Successful extraction of microstructural features from SEM data could be used to support and guide future experimental design thereby reducing the time and cost of technology validation and deployment.

In this work, we consider the task of machine-assisted classification of SEM images on the basis of the temper applied after manufacturing them. We address the challenges of interpretability and explainability using topological representation of SEM images paired with sparse support vector machines. The topological features are benchmarked against conventional convolutional neural networks applied to the SEM images directly. Models are trained on limited data and evaluated on larger collections of images derived from previously unseen material samples to simulate a traditional human-in-the-loop, data-limited analysis scenario. We show that a topological representation outperforms learned features for temper classification, generalizes better to unseen experiments, and is more robust to noise.

%\hjk{(This is purely an observation, but I am a big fan of the new convention of having a section that lists the contributions of this work in bullet points instead of a layout of the paper structure.)}
%The remainder of this paper is organized as follows. In Section \ref{sec:background}, we provide a brief introduction to the Shear Assisted Processing Extrusion (ShAPE) manufacturing technology and the relevant topological tools that support our topological representation. Next, in Section \ref{sec:technical}, we present our technical approach and experimental design. Results are shown in Section \ref{sec:exp} and followed by conclusions and a discussion of future work in Section \ref{sec:disc}.
%Data-driven approaches for reducing the time and financial burden of evaluation experimental design are limited by small data and a preference for explainable, interpretable models for replacing a task traditionally performed and guided by human intuition.
\section{Background}
\label{sec:background}

\subsection{Shear Assisted Processing Extrusion Manufacturing}
\label{subsec:ShAPE}
In this work, we consider AA7075 (aluminum) tubes produced using a novel solid phase processing technique called Shear Assisted Processing Extrusion \cite{WHALEN2021699,shaped1}. During ShAPE manufacturing, a rotating die impinges on a stationary billet housed in an extrusion container with a coaxial mandrel. Due to the shear forces applied on the billet as well as the friction at the tool/billet interface, the temperature increases, and the billet material is plasticized. As the tool traverses into the plasticized material at a predetermined feed rate, the billet material emerges from a hole in the extrusion die to form the tube extrudate.

The ShAPE technology has been able to produce tubes from AA7075 that possess material properties that have been previously unachievable. At present, however, only nascent models exist for simulating the microstructures resulting from ShAPE and any subsequent tempering. As a newly developed manufacturing methodology with emerging first principles models, there are a limited number of ShAPE experiments from which to learn about these fundamental relationships.

AA7075 sample tubes considered in this work are produced using a variety of ShAPE processing parameters including feed rate and die rotation rate (spindle speed). Different sections of the extruded sample are then tempered to T5 and T6 conditions to understand the evolution of as-extruded microstructures and resulting properties as a consequence of the heat treatment. Images are taken, using SEM at a magnification of 500X, of samples for each of the three possible tempered states: ``as extruded" (e.g. pre-temper), T5, and T6. Each sample also corresponds to measurable material properties that will be considered in future work.

\begin{figure}
     \centering
     \begin{subfigure}[b]{.8\textwidth}
            \centering
            \includegraphics[width=0.7\textwidth]{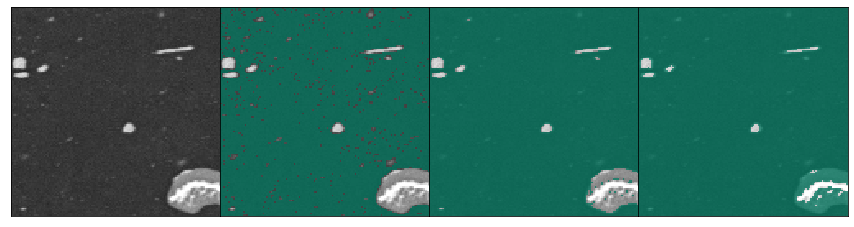}
            \includegraphics[width=0.2\textwidth]{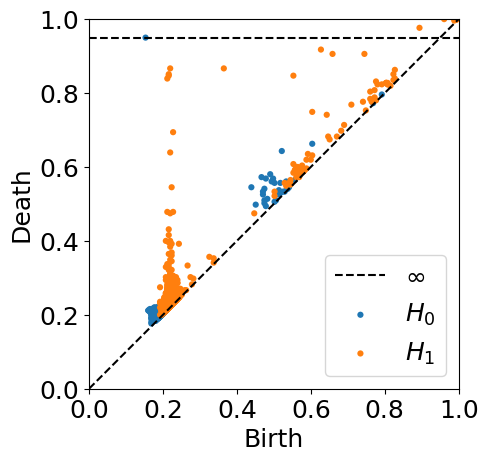}
            \caption{Example image crop for ``as extruded" condition.}
            \label{fig:SEM As Extruded}
     \end{subfigure}
     \\
     \begin{subfigure}[b]{.8\textwidth}
         \centering
         \includegraphics[width=0.7\textwidth]{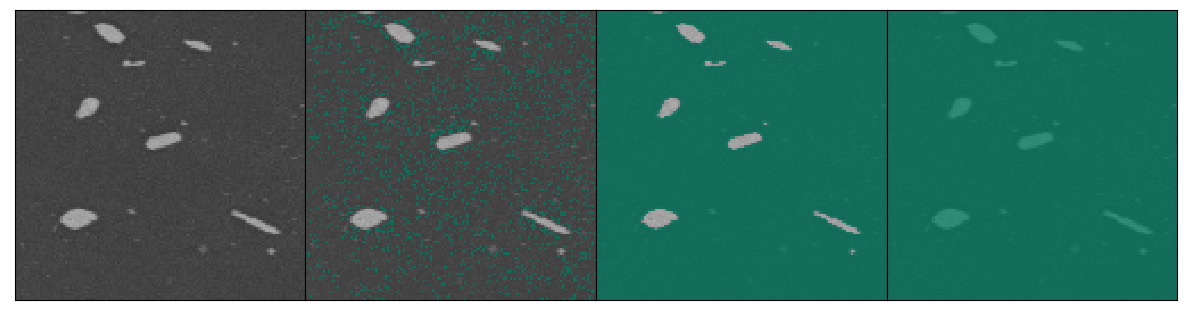}
         \includegraphics[width=0.2\textwidth]{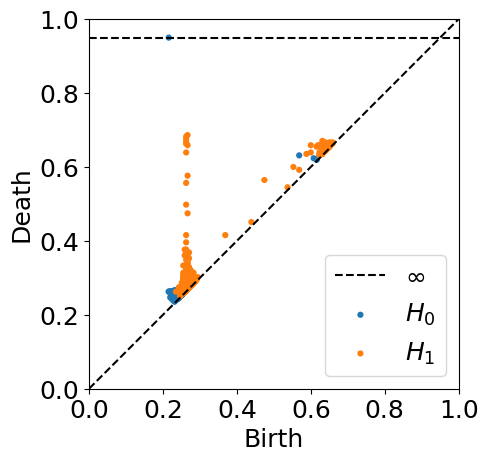}
        \caption{Example image crop for T5 temper condition.}
    \label{fig:SEM T5}
     \end{subfigure}
     \\
     \begin{subfigure}[b]{.8\textwidth}
         \centering
         \includegraphics[width=0.7\textwidth]{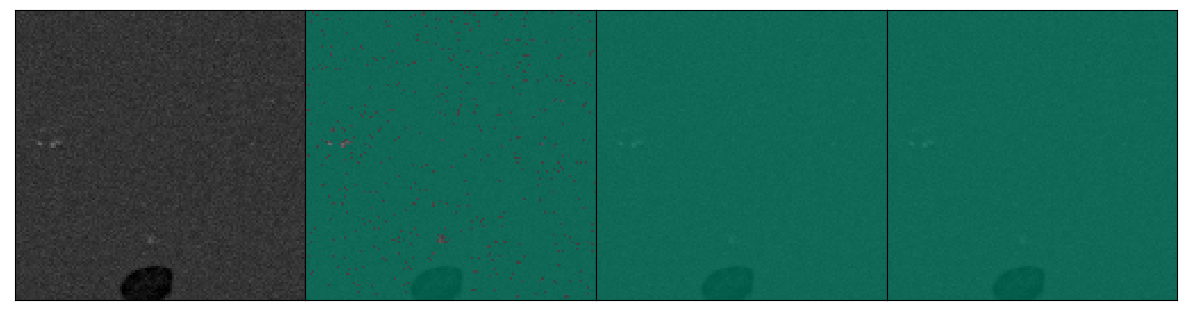}
         \includegraphics[width=0.2\textwidth]{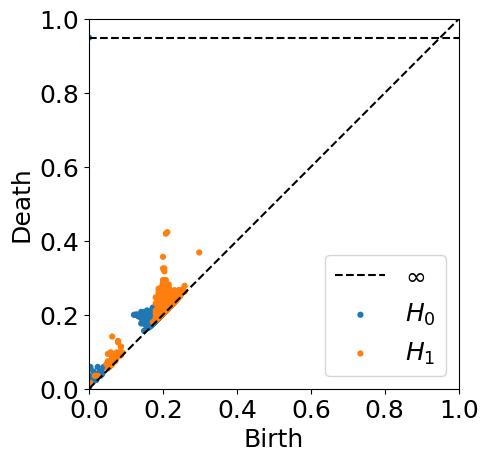}
        \caption{Example image crop for T6 temper condition.}
    \label{fig:SEM T6}
     \end{subfigure}
        \caption{An example image crop for each temper condition (``as extruded", T5, T6), the lowerstar filtations (for threshold values 0 , 0.25, 0.5, and 0.75), and the resulting persistence diagrams.}
        \label{fig:SEM_PD_summary}
\end{figure}

\subsection{Persistent Homology}
\label{subsec:PH}
\emph{Persistent homology} is a popular tool from algebraic topology used to study shapes of datasets \cite{edelsbrunner2000topological}.
It tracks the birth and death of topological features particularly $k$-dimensional holes (connected components, loops, voids, etc.) along a nested sequence of topological spaces.
The $k$-dimensional holes of a topological space $X$ are encoded in an algebraic structure called the $k$-th homology group of $X$ denoted by $H_k(X)$.
Considering a filtration of the spaces, the birth and death of the homological features can be encoded in a \emph{persistence diagram}.

\emph{Lower star filtration} (LSF), or \emph{sublevel set filtration}, is a technique for gridded/pixel data where the nested sequence of topological spaces are constructed based on the function value at each pixel.
It constructs a sparse distance matrix in which every pixel in the image is a vertex, and every vertex is connected to its 8 adjacent neighboring pixels where the distance between
two adjacent pixels is defined to be the maximum of their values \cite{tralie2018ripser}.
Furthermore, a pixel appears in
the filtration only when its value is less than the specified filtration threshold.

\section{Technical Approach and Experimental Design}
\label{sec:technical}
Conversations with material science SMEs revealed that SEM images of AA7075 samples produced using ShAPE manufacturing have multiple significant features such as grain boundary density, precipitate density and morphology as well as void size and distribution. Of these, precipitate morphology and topology are known to be the most affected as a result of the tempering process. In total there are approximately 30 AA7075 tubes manufactured using ShAPE that were considered in this study. These samples came from different combinations of process parameters prior to being tempered. Each experimental sample is subdivided and certain subsections are tempered to T5 or T6 conditions. SEM images were taken for each of the AA7075 tubes that can give rise to numerous smaller, cropped SEM images.
%Conversations with material science SMEs revealed that SEM images of AA7075 samples produced using ShAPE manufacturing are considered based on the relative size and distribution of precipitate grain structure. In total there are approximately 30 samples that have been manufactured using ShAPE. These samples came from different combinations of process parameters prior to being tempered. Each experimental sample is subdivided and certain subsections are tempered to T5 or T6 conditions. Additionally, each subsection of the sample corresponds to a single SEM image that can give rise to numerous smaller, cropped SEM images.

The cropped grayscale SEM images, chosen to be size $128\times128$, can be considered as bounded, 2-dimensional functions and therefore can be summarized by their sublevelset topology as described in Section \ref{subsec:PH}. Based on the characteristics used by SMEs we focus on the 1-dimensional homology of the grayscale images. Example image crops for each temper condition are shown in the leftmost plots of Figure \ref{fig:SEM_PD_summary}. In addition, Figure \ref{fig:SEM_PD_summary} showcases the lower star filtation\footnote{We use Ripser's \cite{tralie2018ripser} implementation of LSF and for persistence images we use Gaussian distributions
%: $\boldsymbol \mu = \begin{bmatrix} 0 \\0 \end{bmatrix}$ $\boldsymbol \Sigma = \begin{bmatrix} 0.003 & 0 \\0 & 0.003 \end{bmatrix}$ 
and weighting functions depending on vertical persistence coordinates. % with n=3
} and resulting persistence diagrams for each example. For each example of the LSF at a given threshold, the green pixels are those which form the simplicial complex at the indicated threshold. A finer scale of filtration thresholds are provided in Appendix \ref{app:thresholding}. 

For each SEM image crop, we create a $10\times 10$ pixel persistence image \cite{adams2017persistence} that is then vectorized and used as the input into a sparse support vector machine (SSVM) classifier. There are three temper classes considered: ``as extruded", T5, and T6. The SSVM classifier is trained using a one versus rest approach resulting in three dividing hyperplanes\footnote{We use Scikit-learn \cite{pedregosa2011scikit} implementation of Linear SVM, \emph{LinearSVC}, with l1-penalty and squared hinge loss. We search over different values of regularization parameter $10^{-3},10^{-2},10^{-1}$, and 1.}. We refer to the process of applying an SSVM classifier to topological features extracted using the LSF to temper-dependent micro-structure as \emph{TopTemp}. As a benchmark, we compare the performance of an SSVM applied to our explainable topological representation to a convolutional neural network\footnote{We fine tune a pre-trained ResNet18 \cite{he2016deep,torchvision}  model on the 50 samples from each temper class.} applied to the cropped SEM images directly. Both models are trained using the same train/test split consisting of 50 random SEM image crops from each temper class for training and 475 random crops from each class for testing. Train/test crops are drawn from disjoint sets of samples to evaluate model generalization. We also create a second test set by adding random Gaussian noise to the original test set with different values of variance to evaluate model robustness. 
%\hjk{(It is probably worth commenting on why your train/test splits are so unusual. That is, explicitly say you are doing this to simulate the low data regime. In a future work it would be interesting to see what would happen if shifted the amount of data. At what point does the ResNet18 catch up? If ever?)} 
\section{Experimental Results}
\label{sec:exp}
\begin{figure}
    \centering
    \includegraphics[height=0.28\textwidth]{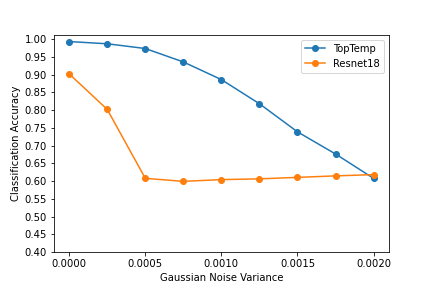} \includegraphics[height=0.25\textwidth]{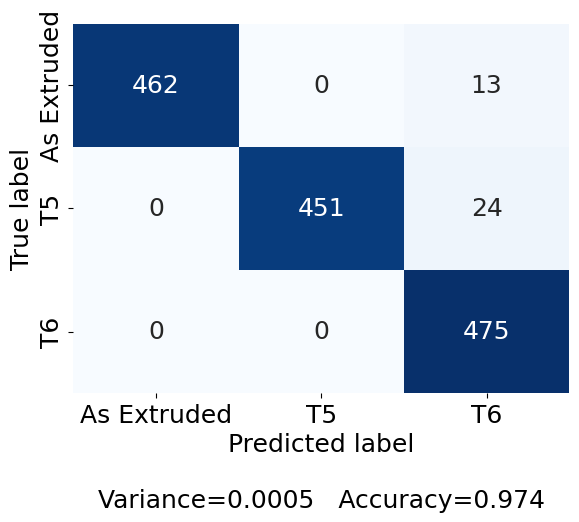}
    \includegraphics[height=0.25\textwidth]{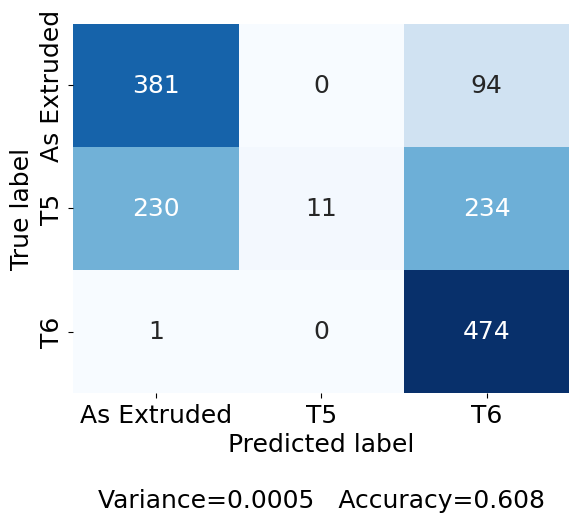}
    \caption{The classification test accuracies for \emph{TopTemp} and ResNet18 for different levels of Gaussian noise are shown in the leftmost plot. The confusion matrices for Gaussian noise variance 0.0005 are shown for \emph{TopTemp} SSVM (middle) and the fine tuned ResNet18 model (right).}
    \label{fig:variance_vs_accuracies}
\end{figure}

Our core results are summarized in Figure \ref{fig:variance_vs_accuracies}. In the leftmost plot, we show the model accuracies as a function of the additive Gaussian noise with mean 0 and different variance values (for pixel scale $[0,1]$) for both the \emph{TopTemp} model and the fine-tuned ResNet18 model. 
Across all noise levels, including noise-free, \emph{TopTemp} outperforms the ResNet18 model. As seen in Figure \ref{fig:variance_vs_accuracies}, the ResNet18 model's performance drops off rapidly and then stabilizes at approximately the same model accuracy as the worst \emph{TopTemp} accuracy. The two confusion matrices to the right of the plot in Figure \ref{fig:variance_vs_accuracies} provide a more detailed perspective on the model performances for the 0.0005 additive Gaussian noise test set where there is the greatest discrepancy in model performance.

All errors from the \emph{TopTemp} model come from misclassification of ``as extruded'' or T5 samples as T6 samples. Alternatively, the ResNet18 model fails by roughly splitting the T5 samples across the other two classes as well as a substantial number of ``as extruded'' samples into T6. As can be seen in the leftmost images of Figure \ref{fig:SEM_PD_summary}, T6 images can present with a structure that strongly resembles Gaussian noise. Therefore, it is reasonable to assume that a model built on features capturing meaningful structure would have a failure mode where ``as extruded'' or T5 samples would be classified as T6 in the presence of additive white noise. This is strongly suggestive that \emph{TopTemp} captures meaningful, interpretable, and more robust features than the ResNet18 model. The pattern is maintained across all noise levels.

In addition to outperforming the ResNet18 model, the SSVM classifier applied to \emph{TopTemp} features yields a second level of explainability. Appendix \ref{app:ssvm} shows reshaped SSVM weight vectors for each of the dividing hyperplanes. From the dividing hyperplane weights it can be seen that the most important regions of the topological representation for differentiating T6 crops from the other classes are those that correspond to low persistence features, e.g. small, low contrast precipitates. Similar relationships, consistent with SME descriptions, can be seen by observing the average topological representations shown in Appendix \ref{app:averages}.
%\begin{figure}
%    \centering
%    \includegraphics[width=0.45\textwidth]{figures/train_confusion.png}
%    \includegraphics[width=0.45\textwidth]{figures/noisy_confusion_0.0.png}\\
%    \includegraphics[width=0.45\textwidth]{figures/noisy_confusion_0.001.png}
%    \includegraphics[width=0.45\textwidth]{figures/noisy_confusion_0.002.png}
%    \caption{The confusion matrices for train set and test sets with various Gaussian noise variance 0.001 and 0.002 (bottom: left, right).}
%    \label{fig:conf_mat}
%\end{figure}

\section{Conclusions and Future Work}
\label{sec:disc}
In this work, we have presented a novel approach for using a topological representation for temper classification for SEM images of AA7075 tube samples produced using the innovative ShAPE manufacturing process. The developed technique, called \emph{TopTemp}, is able to achieve superior temper classification accuracy relative to a conventional deep learning architecture. Additionally, \emph{TopTemp} is able to maintain its superior classification accuracy when tested on a large collection of images corresponding to previously unseen material samples. Finally, \emph{TopTemp} performance is less impacted when evaluated on test images corrupted by noise as observed in Figure \ref{fig:variance_vs_accuracies}. The misclassification that occurs in the presence of noise is explainable and further reflects that topological representations capture domain relevant structure. 

This work provides a valuable first step for demonstrating the ability to parse precipitate structure using domain-relevant, explainable features based on temper topology. Building on this foundation, in future work we explore the utility of topological features for connecting additional process parameters to the material properties of samples they produce. Successful use of topological features for relating process parameters to material properties would provide a robust, generalizable method that could be used to inform and explain future experimental design.

%\subsubsection*{Acknowledgments}
%KSK thanks Scott Whalen, Md. Reza-E-Rabby, Jens Darsell, and Timothy Roosendaal for their insights into AA7075 manufacturing and property determination. KSK is grateful for the discussions on advanced manufacturing with Cindy Powell and Glenn Grant.

%This research was supported by the Mathematics for Artificial Reasoning in Science (MARS) initiative via the Laboratory Directed Research and Development (LDRD) investments at Pacific Northwest National Laboratory (PNNL). PNNL is a multi-program national laboratory operated for the U.S. Department of Energy (DOE) by Battelle Memorial Institute under Contract No. DE-AC05-76RL0-1830.
\bibliography{main}
\bibliographystyle{plain}

\appendix
\section{\emph{TopTemp} Fine-scale Thresholding}
\label{app:thresholding}
\begin{figure}
     \centering
     \begin{subfigure}[b]{.8\textwidth}
            \centering
            \includegraphics[width=0.7\textwidth]{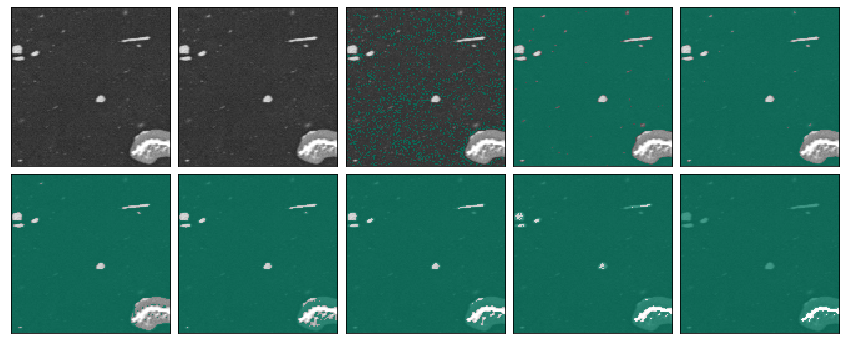}
            \includegraphics[width=0.25\textwidth]{figures/as_extruded_PD.png}
            \caption{Example image crop for ``as extruded" condition.}
            \label{fig:SEM As Extruded}
     \end{subfigure}
     \\
     \begin{subfigure}[b]{.8\textwidth}
         \centering
         \includegraphics[width=0.7\textwidth]{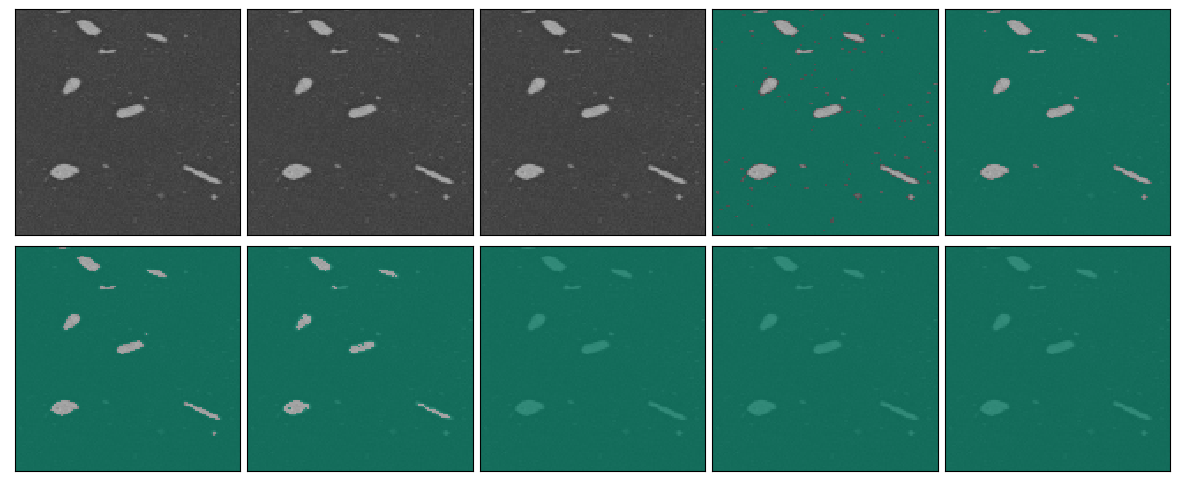}
         \includegraphics[width=0.25\textwidth]{figures/T5_PD.png}
        \caption{Example image crop for T5 temper condition.}
    \label{fig:SEM T5}
     \end{subfigure}
     \\
     \begin{subfigure}[b]{.8\textwidth}
         \centering
         \includegraphics[width=0.7\textwidth]{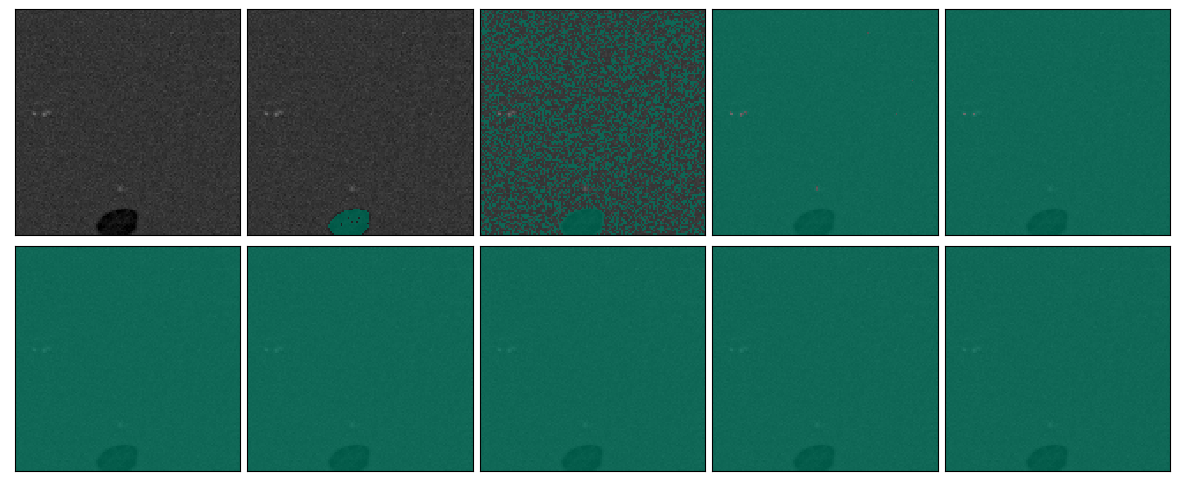}
         \includegraphics[width=0.25\textwidth]{figures/T6_PD.png}
        \caption{Example image crop for T6 temper condition.}
    \label{fig:SEM T6}
     \end{subfigure}
        \caption{An example image crop for each temper condition (``as extruded", T5, T6), the lowerstar filtations (for threshold values $0 , 0.1, 0.2, \cdots, 0.9$), and the resulting persistence diagrams.}
        \label{fig:SEM_PD}
\end{figure}

In Figure \ref{fig:SEM_PD}, we showcase the lower star image filtrations for a finer choice of thresholds.
We observe from the persistence diagrams that the ``as extruded" SEM sample contains persistent 1-dimensional homological features corresponding to high contrast patches 
%(i.e., large variability in pixel intensities) 
compared to the T6 SEM sample with no high contrast patches.
Indeed, the T5 SEM sample lies in the middle of both cases with not-so-high contrast patches.

\section{SSVM Feature Importance}
\label{app:ssvm}
\begin{figure}
    \centering
    \includegraphics[height=3.5cm]{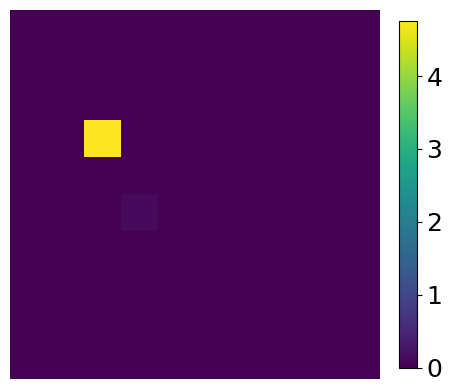}
    \includegraphics[height=3.5cm]{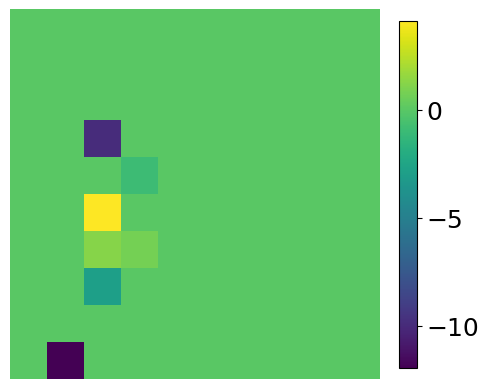}
    \includegraphics[height=3.5cm]{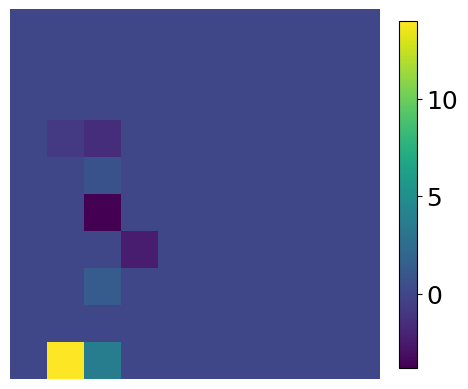}
    \caption{The reshaped SSVM weight vectors for each of the dividing hyperplanes for each class using a one versus rest approach: (left) ``as extruded"; (middle) T5; (right) T6. }
    \label{fig:svm_features}
\end{figure}

In Figure \ref{fig:svm_features}, we showcase the resulting reshaped sparse SSVM weight vectors indicating the most important regions in the persistence images that differentiate each class from the rest.
We observe that persistent 1-dimensional homological features distinguish the ``as extruded" class from the rest.
In contrast, early low-persistence 1-dimensional homological features identify the T6 class from the rest.

\section{Average Persistence Images}
\label{app:averages}
\begin{figure}
    \centering
    \includegraphics[height=3.5cm]{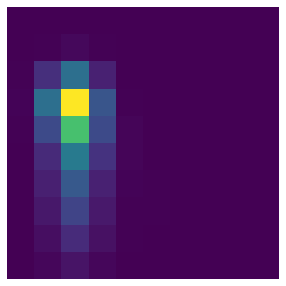}
    \includegraphics[height=3.5cm]{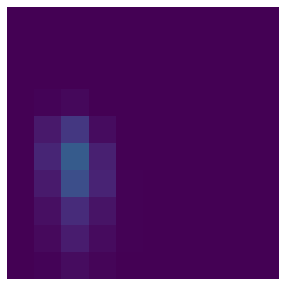}
    \includegraphics[height=3.5cm]{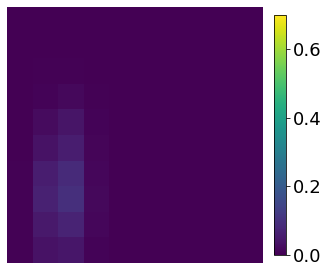}
    \caption{The average $10 \times 10$ persistence image constructed from the peristence diagram of the 1-dimensonal homology for each temper condition: (left) ``as extruded"; (middle) T5; (right) T6.}
    \label{fig:avg PI}
\end{figure}

We showcase in Figure \ref{fig:avg PI}, the average $10\times10$ $H_1$ persistence images generated from the train data for each class.
The average persitence image corresponding to ``as extruded'' condition showcases features with a higher lifetime (vertical persistence coordinate) indicating persistent 1-dimensional homological features compared to the persistence image of T5 temper conditions with less prominent 1-dimensional homological features.

\end{document}

%% file: math_commands.tex
\usepackage{amsmath,amsfonts,bm}

\def\eqref#1{equation~\ref{#1}}
\def\1{\bm{1}}

\DeclareMathAlphabet{\mathsfit}{\encodingdefault}{\sfdefault}{m}{sl}
\SetMathAlphabet{\mathsfit}{bold}{\encodingdefault}{\sfdefault}{bx}{n}